\documentclass[12pt]{article}
\usepackage{subeqn}
\usepackage{epsfig}
\newcommand{\be}{\begin{equation}}
\newcommand{\ee}{\end{equation}}

\newcommand{\no}{\noindent}
\newcommand{\ce}{\begin{center}}
\newcommand{\nc}{\end{center}}

\makeatletter
\@addtoreset{equation}{section}
\makeatother

\def\sqr#1#2{{\vcenter{\vbox{\hrule height.#2pt
 \hbox{\vrule width.#2pt height#1pt \kern#1pt
 \vrule width.#2pt} \hrule height.#2pt}}}}

\def\operp{\hbox{${\kern+.25em{\bigcirc}
\kern-.85em\bot\kern+.85em\kern-.25em}$}}

\def\lsim{\;\raise0.3ex\hbox{$<$\kern-0.75em\raise-1.1ex\hbox{$\sim$}}\;}
\def\gsim{\;\raise0.3ex\hbox{$>$\kern-0.75em\raise-1.1ex\hbox{$\sim$}}\;}
\def\no{\noindent}

\def\ce{\centerline}
\def\ve{\vfill\eject}
\def\rdots{\mathinner{\mkern1mu\raise1pt\vbox{\kern7pt\hbox{.}}\mkern2mu
 \raise4pt\hbox{.}\mkern2mu\raise7pt\hbox{.}\mkern1mu}}

\def\e e{$e^+ e^-$ }

\begin{document}

\ce{\bf KNOTS AND PREONS}

\vskip.3cm

\ce{\it Robert J. Finkelstein}
\vskip.3cm

\ce{Department of Physics and Astronomy}
\ce{University of California, Los Angeles, CA 90095-1547}

\vskip1.0cm

\no{\bf Abstract.}  It is shown that
the four quantum trefoil solitons that are described
by the irreducible representations ${\cal{D}}^{3/2}_{mm^\prime}$
of the quantum algebra $SL_q(2)$ (and that may be identified with
the four families of elementary fermions $(e,\mu,\tau;\nu_e
\nu_\mu\nu_\tau;d,s,b;u,c,t)$) may be built out of three preons,
chosen from two charged preons with charges (1/3,-1/3) and two
neutral preons.  These preons are fermions and are
described by the ${\cal{D}}^{1/2}_{mm^\prime}$, representation
of $SL_q(2)$.  There are also four bosonic preons described by
the ${\cal{D}}^1_{mm^\prime}$ and ${\cal{D}}^0_{00}$ 
representations of $SL_q(2)$.  The knotted standard theory may be
replicated at the preon level and the conjectured particles are
in principle indirectly observable.

\ve

\section{Introduction}

One proposal for going beyond the standard theory of elementary
particles depends on the use of the quantum groups.  A particular
form of this proposal, which exploits the fact that the quantum
group $SL_q(2)$ is the symmetry group of the knot, has revived
interest in the old speculation that the elementary particles
are knots.$^{1,2,3}$  This possibility may be empirically tested by
comparing the four simplest quantum knots 
(quantum trefoils) with the four
classes of simplest particles (neutrinos, leptons, up quarks,
down quarks), and is in fact supported by simple relations that
exist between the topology of the trefoil and the charge,
hypercharge, and the isotopic structure of the simplest particles.
These relations depend on the hypothesis that the kinematic
eigenstates of the quantum knot are matrix elements of
irreducible representations of
$SL_q(2)$, denoted by ${\cal{D}}^{N/2}_{\frac{w}{2}\frac{r+1}
{2}}$, where $(N,w,r)$ are (the number of crossings, writhe,
and rotation of the knot), respectively.  For trefoils one has
$(N=3,w=\pm 3,r=\pm 2)$.  The four quantum trefoils may in this
representation be uniquely correlated with the four classes of
fermions.  The three fermions in each class (e.g. $e,\mu,\tau$ in
the lepton class) are assumed to be three quantum states of a
soliton with the topology of a trefoil.

In a classification where ${\cal{D}}^{N/2}_{\frac{w}{2}
\frac{r+1}{2}}$ represents a bosonic knot 
when $N$ is even, and a fermionic knot
when $N$ is odd, the $N=3$ trefoils represent the simplest 
classes of fermions.  If, however, the new idea that is being
proposed to supplement the standard theory is not primarily the
idea of knots but rather the introduction of a new symmetry, namely
$SL_q(2)$, then the lower representations 
${\cal{D}}^j_{mm^\prime}$ when $j=1/2$ and $j=1$, become of
interest.  We are here interested in relating these ``preon"
representations to the trefoil and other representations.

\ve

\section{The Knot Representation}

We replace the point particles of standard theory with quantum
knots by attaching to each normal mode a knot state just as one
introduces spin by attaching a spin state.  The knot states and
the corresponding fields are defined only up to a gauge
transformation and the action is required to be invariant under
these gauge transformations.

We assume that the kinematical quantum states of the knot are
derived from the 
irreducible representations of the symmetry algebra of the knot,
namely$^3$
\be
{\cal{D}}^j_{mm^\prime}(a,b,c,d) = \sum_{\scriptstyle s\leq n_+
\atop\scriptstyle t\leq n_-} 
{\cal{A}}^j_{mm^\prime}(q,s,t)\delta(s+t,n_+^\prime)
a^sb^{n_+-s}c^td^{n_--t}
\ee
where
\begin{eqnarray}
n_\pm &=& j\pm m \\
n_\pm^\prime &=& j\pm m^\prime
\end{eqnarray}
and the arguments of ${\cal{D}}^j_{mm^\prime}$ obey the algebra
of $SL_q(2)$ as follows:
\be
\left.
\begin{array}{l}
ab = qba \\ ac = qca
\end{array} \right. \quad \left.
\begin{array}{l}
bd = qdb \\ cd = qdc
\end{array} \right. \quad \left.
\begin{array}{l}
bc = cb \\ \hfil 
\end{array} \right. \quad \left.
\begin{array}{l}
ad-qbc=1 \\ da-q_1cb=1 
\end{array} \right.  \qquad
q_1=q^{-1}
\ee

We shall refer to (2.4) as the ``knot algebra".  This algebra and
${\cal{D}}^j_{mm^\prime}(a,b,c,d)$ are defined only up to the
following gauge transformation
\be
\left.
\begin{array}{l}
a^\prime = e^{i\varphi_a}a \\ d^\prime = e^{-i\varphi_a}d
\end{array} \right. \qquad \left.
\begin{array}{l}
b^\prime = e^{i\varphi_b}b \\ c^\prime = e^{-i\varphi_b}c
\end{array} \right.
\ee
Eqs. (2.5) leave the algebra (2.4)
invariant and induce on the elements
of every representation the following transformation$^3$
\be
\begin{array}{rcl}
{\cal{D}}^j_{mm^\prime}(a^\prime,b^\prime,c^\prime,d^\prime)
&=& e^{i(m+m^\prime)\varphi_a}e^{i(m-m^\prime)\varphi_b}
{\cal{D}}^j_{mm^\prime}(a,b,c,d) \\
&=& e^{im(\varphi_a+\varphi_b)}e^{im^\prime(\varphi_a-\varphi_b)}
{\cal{D}}^j_{mm^\prime}(a,b,c,d)
\end{array}
\ee
We now associate ${\cal{D}}^j_{mm^\prime}$ with the geometrical
knot by setting
\be
\begin{array}{rcl}
j &=& \frac{N}{2} \\
m &=& \frac{w}{2} \\
m^\prime &=& \frac{r+1}{2}
\end{array}
\ee
where $w$ and $r$ are the writhe and rotation respectively and
where $N$ is the number of crossings.
The writhe and the rotation are topological invariants and we
shall also assume that $N$ is a dynamical invariant.  
The kinematic states of the quantum knot may be labelled by 
these three
invariants of the motion.  Then by (2.6) and (2.7)
\be
\begin{array}{rcl}
{\cal{D}}^{N/2}_{\frac{w}{2}\frac{r+1}{2}}
(a^\prime,b^\prime,c^\prime,d^\prime)&=& e^{i\frac{w}{2}
\varphi_w} e^{i\frac{r+1}{2}\varphi_r}
{\cal{D}}^{N/2}_{\frac{w}{2}\frac{r+2}{2}}(a,b,c,d) \\
&=& e^{-\frac{i}{k}Q(w)\varphi_w}
e^{-\frac{i}{k}Q(r)\varphi_r}
{\cal{D}}^{N/2}_{\frac{w}{2}\frac{r+1}{2}}(a,b,c,d)
\end{array}
\ee
where
\begin{eqnarray}
Q(w) &=&-k\frac{w}{2} \\
Q(r) &=& -k\frac{r+1}{2}
\end{eqnarray}
and $k$ is a constant to be determined.  $Q(w)$ and $Q(r)$ are two
topological
integrals of the motion.  The gauge transformations (2.8)
operate on all the normal modes and therefore on all fields.
We require that the 
action be invariant under these gauge transformations since
they are induced by the transformations (2.5) that leave the
defining
algebra (2.4) invariant.  Therefore 
by Noether's theorem $Q(w)$ and $Q(r)$ behave as
conserved charges and will be called the writhe charge and the
rotation charge.

We shall now make a direct comparison between $Q(w)$ and $Q(r)$ 
of the quantum knot and the charge and hypercharge of the 4
fermion families, each denoted by $(f_1,f_2,f_3)$.$^3$

\begin{table}[h]
\begin{center}
\begin{tabular}{ccccc|ccccc}
\multicolumn{5}{c}{\underline{Standard Representation}} &
\multicolumn{5}{c}{\underline{Knot Representation}} \\
\underline{$(f_1,f_2,f_3)$} & \underline{$t$} & 
\underline{$t_3$} &
\underline{$t_0$} & \underline{$Q_e$} & \underline{$(w,r)$} &
\underline{${\cal{D}}^{N/2}_{\frac{w}{2}\frac{r+1}{2}}$} &
\underline{$Q_w$} & \underline{$Q_r$} & \underline{$Q_w+Q_r$} \\
$(e,\mu,\tau)$ & $\frac{1}{2}$ & $-\frac{1}{2}$ & $-\frac{1}{2}$
& $-e$ & (3,2) & ${\cal{D}}^{3/2}_{\frac{3}{2}\frac{3}{2}}$ &
$-k\left(\frac{3}{2}\right)$ & $-k\left(\frac{3}{2}\right)$ &
$-3k$ \\
$(\nu_e,\nu_\mu,\nu_\tau)$ & $\frac{1}{2}$ & $\frac{1}{2}$ &
$-\frac{1}{2}$ & 0 & (-3,2) & ${\cal{D}}^{3/2}_{-\frac{3}{2}
\frac{3}{2}}$ & $-k\left(-\frac{3}{2}\right)$ &
$-k\left(\frac{3}{2}\right)$ & 0 \\
$(d,s,b)$ & $\frac{1}{2}$ & $-\frac{1}{2}$ & $\frac{1}{6}$ &
$-\frac{1}{3}e$ & (3,-2) & ${\cal{D}}^{3/2}_{\frac{3}{2}-\frac
{1}{2}}$ & $-k\left(\frac{3}{2}\right)$ & $-k\left(-\frac{1}{2}
\right)$ & $-k$ \\ 
$(u,c,t)$ & $\frac{1}{2}$ & $\frac{1}{2}$ & $\frac{1}{6}$ &
$\frac{2}{3}e$ & (-3,-2) & ${\cal{D}}^{3/2}_{-\frac{3}{2}
-\frac{1}{2}}$ & $-k\left(-\frac{3}{2}\right)$ &
$-k\left(-\frac{1}{2}\right)$ & $2k$ \\
\end{tabular}
\end{center}
\end{table}
\begin{center}
{\bf Table 1.}
\end{center}

\no In this table $Q_w$ and $Q_r$ are given by (2.9) and 
(2.10), and we have adopted a particular relative order of
the four trefoils with respect to the four families of fermions.
For this order we find
\begin{eqnarray}
Q_w &=& et_3 \\
Q_r &=& et_0 \\
Q_w + Q_r &=& Q_e
\end{eqnarray}
\underline{if and only if $k=e/3$}.  These relations are in
agreement with the following independent equation of standard theory:
\be
Q_e = e(t_3+t_0)
\ee
If one aligns the trefoils and the families in any other relative
order, one needs more than one value of $k$.  In this respect
the correspondence between the two sets is unique. 

The correspondence between the quantum knots and the isospin and
charge of the elementary fermions may be summarized by the
following relations, which may also be read directly from Table 1:
\be
\left.
\begin{array}{l}
t~=\frac{N}{6} \\ t_3=-\frac{w}{6} \\ t_0=-\frac{r+1}{6} \\
Q_e=-\frac{e}{6}~(w+r+1) 
\end{array} \right. \qquad \mbox{or}  \qquad \left.
\begin{array}{l}
t~=\frac{j}{3} \\ t_3=-\frac{m}{3} \\ t_0=-\frac{m^\prime}{3} \\
Q_e=-\frac{e}{3}~(m+m^\prime)
\end{array} \right.
\ee
Note also that
\be
Q_e=-\frac{e}{N}\left(\frac{w+r+1}{2}\right) \quad \mbox{and}
\quad Q_e = -e\left(\frac{m+m^\prime}{2j}\right)
\ee
holds for all the elementary fermions.

While $t_3$ and $t_0$ are initially defined in the standard
theory with respect to $SU(2)\times U(1)$, here they label the
gauge transformations on the knot algebra, and are also
described by $(w,r)$ and $(m,m^\prime)$ as shown in Table 1 and
Eq. (2.15).  In the limit of the standard theory, where
$q=1$, $t_3$ and $t_0$ assume their usual meaning in
$SU(2)$ and $U(1)$ respectively.

The states labelled by the quantum numbers $(j,m,m^\prime)$ or
$(t,t_3,t_0)$ are the multinomials lying in the algebra (2.4) and
given by (2.1).  These multinomials are associated with the knot
$(N,w,r)$ and, like the Jones (Laurent) polynomial, label the
knot.

These conclusions may be summarized by assigning ${\cal{D}}^{3t}_{-3t_3-3t_0}$ to the fermion families with $(t,t_3,t_0)$.
The members of each family are described by the quantum states
${\cal{D}}^{3t}_{-3t_3-3t_0}|n\rangle$ where $|n\rangle$ are the
states defined by the knot algebra.$^3$

\no {\it Remark:}  Although the ${\cal{D}}^{3/2}_{\frac{w}{2}
\frac{r+1}{2}}$ representation permits a unified description
of the electroweak properties of the four families, it does 
\underline{not} also imply that they transform 
under a 4-dimensional
representation of $SL_q(2)$.  Rather Eq. (2.8) is the only
formal property of ${\cal{D}}^{N/2}_{\frac{w}{2}\frac{r+1}{2}}
(a,b,c,d)$ that has been invoked.  The only new invariance 
beyond the standard theory that
has been introduced is invariance under the global gauge
transformations of the knot algebra that are expressed by (2.5)
and (2.6).

\ve

\section{The Preon Representation}

Ignoring numerical normalization one finds by (2.1) that the
four fermionic knots are represented by four monomials in the knot
algebra according to Table 2:
\[
\begin{array}{ccccc}
\underline{\mbox{Solitons}} &
\underline{{\cal{D}}^{N/2}_{\frac{w}{2}\frac{r+1}{2}}} &
\underline{{\cal{D}}^{3/2}_{\frac{w}{2}\frac{r+1}{2}}} &
\underline{Q_e} & \underline{et_0}  \\
(e^-,\mu^-,\tau^-) &
{\cal{D}}^{3/2}_{\frac{3}{2}\frac{3}{2}} & a^3 & -e &
-\frac{e}{2} \\
(\nu_e,\nu_\mu,\nu_\tau) &
{\cal{D}}^{3/2}_{-\frac{3}{2}\frac{3}{2}} & c^3 & 0 
& -\frac{e}{2} \\
(d,s,b) &
{\cal{D}}^{3/2}_{\frac{3}{2}-\frac{1}{2}} & \sim ab^2 &
-\frac{1}{3} e & \frac{e}{6} \\
(u,c,t) &
{\cal{D}}^{3/2}_{-\frac{3}{2}-\frac{1}{2}} & \sim cd^2 &
\frac{2}{3} e  & \frac{e}{6} \\
\end{array}
\]
\begin{center}
{\bf Table 2.}
\end{center}

\no This table may be 
re-interpreted by regarding the element $a$ as
a creation operator for a preon of charge -$e/3$, and hypercharge
$-e/6$ and by
regarding $d$ as a creation operator for a preon of charge
+$e/3$ and hypercharge +$e/6$
while $b$ and $c$ are regarded as creation operators for
neutral preons with hypercharge $e/6$ and -$e/6$ respectively.  This interpretation is consistent with our conclusion from
earlier work$^{2,3}$
that adjoint operators $(a,d)$ correspond to opposite charges and
that the $(b,c)$ sector describes neutral states.  According to the same picture the fermion knots, like the nucleons, are
composed of three fermions, which are now preons.

While the ${\cal{D}}^{3/2}$ representation describes 
the standard elementary
fermions, the ${\cal{D}}^{1/2}$ representations describes preons
as follows:
\[
{\cal{D}}^{1/2}: \qquad  
\begin{array}{c|cc}
{}_m\backslash{}^{m'} & \frac{1}{2} & -\frac{1}{2} \\ \hline 
\frac{1}{2} & a & b \\
-\frac{1}{2} & c & d
\end{array}
\]
If one assigns $(t_3,t_0,Q)$ to the preons by the same rules
that are given in (2.14) and
have been validated for the standard fermions, one has
\[
t_3 = -\frac{m}{3}~, \quad t_0 = -\frac{m^\prime}{3} \quad
\mbox{and} \quad Q = e(t_3+t_0)
\]
Then we find Table 3
\[
\begin{array}{c|ccc}
 & \underline{t_3} & \underline{t_0} & \underline{Q}  \\ 
a & -\frac{1}{6} & -\frac{1}{6} & -\frac{e}{3} \\
c & \frac{1}{6} & -\frac{1}{6} & 0 \\
d & \frac{1}{6} & \frac{1}{6} & \frac{e}{3} \\
b & -\frac{1}{6} & \frac{1}{6} & 0
\end{array}
\]
\begin{center}
{\bf Table 3.}
\end{center}

\no Note that the charge assignments of Table 3 are consistent
with the preon interpretation of Table 2.

\no On the other hand $N=1$ if we maintain $j = \frac{N}{2}$.
Therefore the preon representation as a representation of the
quantum algebra ${\cal{D}}^{1/2}_{mm^\prime}$, does not
correspond to a knot.  It may be pictured as a twisted loop.
If we maintain ${\cal{D}}^{3t}_{-3t_3.-3t_0}$, then $t = 1/6$.

Let us refer to the ${\cal{D}}^{1/2}_{mm^\prime}$ particles as
the fermionic preons and to the ${\cal{D}}^1_{mm^\prime}$
particles as the bosonic preons.  We will extend the definition
of $(t,t_3,t_0)$ to all representations by maintaining the rules
for labelling by isotopic labels as follows:
\be
\left.
\begin{array}{l}
m=-3t_3 \\ m^\prime=-3t_0
\end{array} \right. \quad \left.
\begin{array}{l}
j=3t \\ Q=e(t_3+t_0) 
\end{array} \right.
\ee
and
\be
\begin{array}{rcl}
w &=& -6t_3 \\
r+1 &=& -6t_0
\end{array}
\ee
Now $j=1$ implies $N=2$ and $t=\frac{1}{3}$.  Then 
${\cal{D}}^1_{mm^\prime}$ represents a ``dipreon" with 2
crossings.
For ${\cal{D}}^1_{mm^\prime}$ we find Table 4 (again
ignoring normalizing numerical factors)

\begin{table}[h]
\begin{center}
\begin{tabular}{l|cccc||l|cccc||l|cccc}
& \underline{$t_3$} & \underline{$t_0$} & \underline{$Q/e$} & 
\underline{${\cal{D}}^1_{mm^\prime}$} & &
\underline{$t_3$} & \underline{$t_0$} & \underline{$Q/e$} &
\underline{${\cal{D}}^1_{mm^\prime}$} & &
\underline{$t_3$} & \underline{$t_0$} & \underline{$Q/e$} &
\underline{${\cal{D}}^1_{mm^\prime}$} \\
${\cal{D}}^1_{11}$ & $-\frac{1}{3}$ & $-\frac{1}{3}$ &
$-\frac{2}{3}$ & $a^2$ & ${\cal{D}}^1_{01}$ & 0 &
$-\frac{1}{3}$ & $-\frac{1}{3}$ & $ac$ & 
${\cal{D}}^1_{-11}$ & $\frac{1}{3}$ & $-\frac{1}{3}$ & 0 &
$c^2$ \\
${\cal{D}}^1_{10}$ & $-\frac{1}{3}$ & 0 & $-\frac{1}{3}$ &
$ab$ & ${\cal{D}}^1_{00}$ & 0 & 0 & 0 & $ad+bc$ &
${\cal{D}}^1_{-10}$ & $\frac{1}{3}$ & 0 & $\frac{1}{3}$ &
$cd$ \\
${\cal{D}}^1_{1-1}$ & $-\frac{1}{3}$ & $\frac{1}{3}$ & 0 &
$b^2$ & ${\cal{D}}^1_{0-1}$ & 0 & $\frac{1}{3}$ &
$\frac{1}{3}$ & $bd$ & ${\cal{D}}^1_{-1-1}$ & $\frac{1}{3}$
& $\frac{1}{3}$ & $\frac{2}{3}$ & $d^2$
\end{tabular}
\end{center}
\end{table}
\begin{center}
{\bf Table 4.}
\end{center}

Again in Table 4 $a$ and $d$ can be interpreted as creation 
operators for particles of charge $-e/3$ and $e/3$ respectively,
while $b$ and $c$ are creation operators for neutral particles.

To show that the interpretation of $(a,b,c,d)$ as creation
operators for preons holds in all representations let us introduce
the operators for the writhe, ${\cal{W}}$, the rotation
${\cal{R}}$, and the charge ${\cal{Q}}$, as follows:$^3$
\begin{eqnarray}
\frac{1}{2} {\cal{W}}~{\cal{D}}^j_{mm^\prime} &=& m~
{\cal{D}}^j_{mm^\prime} \\
\frac{1}{2} {\cal{R}}~{\cal{D}}^j_{mm^\prime} &=& m^\prime~
{\cal{D}}^j_{mm^\prime} \\
{}~{\cal{Q}}~{\cal{D}}^j_{mm^\prime} &=& -\frac{e}{3} \frac{{\cal{W}}+{\cal{R}}}{2}~{\cal{D}}^j_{mm^\prime}
\end{eqnarray}
where
\begin{eqnarray}
{\cal{W}} &=& \omega_a-\omega_d+\omega_b-\omega_c \\
{\cal{R}} &=& \omega_a-\omega_d-\omega_b+\omega_c \\
{\cal{Q}} &=& -\frac{e}{3} (\omega_a-\omega_d)
\end{eqnarray}
Here the $\omega_x$ are defined by their action on every term of
${\cal{D}}_{mm^\prime}$ according to
\be
\omega_x(\ldots x^{n_x}\ldots) = n_x(\ldots x^{n_x}\ldots)
\ee
Then by (3.6), (3.9), and (2.1)
\be
{\cal{W}}~{\cal{D}}^j_{mm^\prime} = (n_a-n_d+n_b-n_c)
{\cal{D}}^j_{mm^\prime} 
\ee
and by (3.3)
\be
n_a-n_d+n_b-n_c = 2m
\ee
By (3.7) and (3.9) and (2.1)
\be
{\cal{R}}~{\cal{D}}^j_{mm^\prime} = (n_a-n_d-n_b+n_c)
{\cal{D}}^j_{mm^\prime}
\ee
and by (3.4)
\be
n_a-n_d-n_b+n_c = 2m^\prime
\ee
Eqs. (3.10) and (3.12) depend on the fact that 
$n_a-n_d=m+m^\prime$ and
$n_b-n_c=m-m^\prime$ 
are the same for every term of (2.1).$^3$  By (3.1) and (3.11)
\be
t_3 = -\frac{1}{6}~(n_a-n_d+n_b-n_c)
\ee
By (3.1) and (3.13)
\be
t_0 = -\frac{1}{6}~(n_a-n_d-n_b+n_c)
\ee
Then
\be
\begin{array}{rcl}
Q &=& e(t_3+t_0) \\
&=& \frac{e}{3}~(n_d-n_a)
\end{array}
\ee
Eqs. (3.14), (3.15), and (3.16) hold in every representation and
agree with the interpretation of $a$ and $d$ as creation operators
for preons of charge $-e/3$ and $+e/3$ respectively.  According to
(3.15) and (3.14) all four of the preons contribute
$\pm \frac{1}{6}$ to the hypercharge as well as to $t_3$ in
agreement with the Tables 2, 3, 4.

The operator for $N$ $(=2j)$ is
\[
{\cal{N}} = \omega_a+\omega_b+\omega_c+\omega_d
\]
 and the eigenvalues of ${\cal{N}}$ are
\[
N = n_a+n_b+n_c+n_d
\]
{\bf
Therefore the total number of preons in the knot described by
${\cal{D}}^{N/2}_{\frac{w}{2}\frac{r+1}{2}}$ is equal to the
number of crossings $(N)$}.  If we assume that the preons are
fermions, then a knot with $N$ crossings is a boson or fermion
depending on the parity of $N$, as we have previously assumed.

In this dual description of the knotted soliton the number of
crossings, the writhe and the rotation are dual to the number of
preons, $t_3$ and $t_0$ respectively.  The particle-knot duality
resembles the particle-wave duality, since on the particle-knot
side one has by (3.2)
\[
\begin{array}{rcl}
t_3 &=& -\frac{w}{6} \\
t_0 &=& -\frac{r+1}{6}
\end{array}
\]
while on the particle-wave side one has
\[
p_\mu = \hbar k_\mu
\]
and neither the knot nor the wave can be localized.

Let us also apply the knot relations to the $j=1/2$ and $j=1$
representations.  By (2.7) we have
$N=1$ for $j=1/2$, and $N=2$ for $j=1$.  Although these
are not knots, but twisted loops, $w$ and $r$ still have meaning
and may be computed according to (2.7).  
Although the writhe and rotation are no
longer topologically conserved, for these cases $w$ and $r$
will remain constants of the motion if the Hamiltonian, 
${\cal{H}}$, is so chosen that
\be
\begin{array}{rcl}
\left[{\cal{H}},{\cal{W}}\right] &=& 0 \\
\left[{\cal{H}},{\cal{R}}\right] &=& 0
\end{array}
\ee
In particular $w$ and $r$ are still constants of the motion if
${\cal{H}}$ is a function of $bc$.

\ve

\section{The Intepretation of $SU_q(2)$ Preons}

The preceding discussion is based on $SL_q(2)$.  One may,
however, describe the knots by $SU_q(2)$ instead of by
$SL_q(2)$ if one sets
\be
\begin{array}{rcl}
\bar a &=& d \\ \bar b &=& -qc
\end{array}
\ee
Then the $SU_q(2)$ algebra is
\be
\left.
\begin{array}{l}
ab=qba \\ a\bar b=q\bar ba 
\end{array} \right. \quad \left.
\begin{array}{l}
a\bar a+b\bar b=1 \\ \bar aa+q_1^2\bar bb=1
\end{array} \right. \quad
b\bar b=\bar bb
\ee
The gauge invariance of the $SU_q(2)$ algebra is expressed by
\be
\left.
\begin{array}{l}
a^\prime = e^{i\varphi_a}a \\ \bar a^\prime=e^{-i\varphi_a}\bar a
\end{array} \right. \qquad \left.
\begin{array}{l}
b^\prime = e^{i\varphi_b}b \\
\bar b^\prime = e^{-i\varphi_b}\bar b
\end{array} \right.
\ee
The important relation (2.6) holds exactly if ${\cal{D}}^j_{mm^\prime}$ refers to $SU_q(2)$ as well as to $SL_q(2)$.  One
passes from the representations of $SL_q(2)$ to those of 
$SU_q(2)$ by (4.1). In the $SU_q(2)$ language 
$a$ and $d$ are not creation operators
for distinct preons but are creation operators for preons and
antipreons.  Similarly $b$ and $c$ are related as antiparticles.
Then all the solitons in Table 2 are constructed out of $a$ and
$b$ particles and their antiparticles.

Ref. (3) is written in the $SU_q(2)$ language, but most of 
this paper is written in $SL_q(2)$ language.

\ve

\section{Preons as Physical Particles}

We have so far viewed the preons only as a simple way to describe
the algebraic structure of the knot polynomials.  If these preons
are in fact physical particles, the following decay modes of the
quarks are possible
\[
\begin{array}{llll}
{\rm down~quarks:} & {\cal{D}}^{3/2}_{\frac{3}{2}-\frac{1}{2}}
\longrightarrow {\cal{D}}^{1/2}_{\frac{1}{2}\frac{1}{2}} +
{\cal{D}}^1_{1-1} & & ab^2 \longrightarrow a+b^2 \\
& & {\rm or} & \\
{\rm up~quarks:} & {\cal{D}}^{3/2}_{-\frac{3}{2}-\frac{1}{2}}
\longrightarrow {\cal{D}}^{1/2}_{-\frac{1}{2}\frac{1}{2}} +
{\cal{D}}^1_{-1-1} & & cd^2 \longrightarrow c+d^2
\end{array}
\]
and the preons could play an intermediary role as virtual
particles in quark processes.  

The justification for considering
the preons seriously as physical particles would then no longer 
depend  exclusively on the knot
conjecture but rather on a more general role of $SL_q(2)$ gauge
invariance.  Then the preons would appear as matrix elements of
the fundamental and adjoint representations of $SL_q(2)$ just as
the fermionic and bosonic knots appear in the $j=3/2$ and $j=3$
representations of $SL_q(2)$.$^3$  In this scenario knots would be
just one of the manifestations of a $SL_q(2)$ 
related symmetry.
There would also be no need to introduce a new Lagrangian for the
preons since all particles described by
representations of $SL_q(2)$ would be subject to
the same modified standard action as follows:

\vskip.5cm

In the knotted standard theory at the level of the quarks and 
leptons we attach elements of the ${\cal{D}}^{3/2}_{mm^\prime}$
representation to the standard
quark and lepton normal modes, and 
elements of the ${\cal{D}}^3_{mm^\prime}$ representation to the
normal modes of the standard
vector boson fields.  At the preon level one
attaches elements of the fundamental ${\cal{D}}^{1/2}_{mm^\prime}$
representation to the spinor fields and elements of the adjoint
${\cal{D}}^1_{mm^\prime}$ representation to the vector fields.
 
\vskip.5cm

The simple knot model predicts an unlimited number of excited
states$^{2,3}$ but it appears that there are only three 
generations, e.g.
$(d,s,b)$.  According to the preon scenario, however, it may be
possible to avoid this problem by showing
that the quarks will dissociate into preons if
given a critical ``dissociation energy" less than that needed to
reach the level of the fourth predicted flavor.  In that case one
would also
expect the formation of a preon-quark plasma at sufficiently
high temperatures.  It may be possible to study the thermodynamics of the  plasma composed of quarks and these hypothetical particles.

Since the $a$ and $\bar a$ particles are charged $(\pm e/3)$
one should expect their electroproduction according to
\[
e^++e^- \rightarrow a+\bar a + \ldots
\]
at sufficiently high energies of a colliding $(e^+,e^-)$ pair.

\vskip.5cm

\no {\bf References.}

\begin{enumerate}
\item R. J. Finkelstein, Int. J. Mod. Phys. A{\bf 20}, 6481 
(2005).
\item A. C. Cadavid and R. J. Finkelstein, {\it ibid.}
A{\bf 25}, 4264 (2006).
\item R. J. Finkelstein, {\it ibid.} A{\bf 22}, 4467 (2007).
\end{enumerate}

\end{document}